\documentclass[10pt, conference, letterpaper]{IEEEtran}
\usepackage{cite}
\usepackage{amsmath,amssymb,amsfonts}
\usepackage{algorithmic}
\usepackage{graphicx}
\usepackage{textcomp}
\usepackage{xcolor}
\usepackage{wasysym}
\usepackage{enumitem}
\usepackage{multirow}
\usepackage{lscape}
\def\BibTeX{{\rm B\kern-.05em{\sc i\kern-.025em b}\kern-.08em
    T\kern-.1667em\lower.7ex\hbox{E}\kern-.125emX}}
\begin{document}

\title{Demystifying content-blockers: A large scale study of actual performance gains\\
}

\author{\IEEEauthorblockN{Ismael Castell-Uroz}
\IEEEauthorblockA{\textit{Universitat Polit\`ecnica de Catalunya} \\
Barcelona, Spain \\
icastell@ac.upc.edu}
\and
\IEEEauthorblockN{Josep Sol\'e-Pareta}
\IEEEauthorblockA{\textit{Universitat Polit\`ecnica de Catalunya} \\
Barcelona, Spain \\
pareta@ac.upc.edu}
\and
\IEEEauthorblockN{Pere Barlet-Ros}
\IEEEauthorblockA{\textit{Universitat Polit\`ecnica de Catalunya} \\
Barcelona, Spain \\
pbarlet@ac.upc.edu}
}

\maketitle

\begin{abstract}
With the evolution of the online advertisement and tracking ecosystem, content-filtering has become the reference tool for improving the security, privacy and browsing experience when surfing the Internet. It is also commonly believed that using content-blockers to stop unsolicited content decreases the time needed for loading websites.
In this work, we perform a large scale study with the 100K most popular websites on the actual performance improvements of using content-blockers. We focus our study in the two most relevant metrics for user experience; bandwidth and latency. Our results show that using such tools results in small improvements in terms of bandwidth usage but, contrary to popular belief, it has a negligible impact in terms of loading time. We also find that, in the case of small and fast loading websites, the use of content-blockers can even result in increased browsing latency.
\end{abstract}


\section{Introduction}
The use of content-filtering tools has seen an exponential increase since the initial development of AdBlock Plus \cite{adblock_plus} in 2005, one of the most popular and widely used adblockers. This was one of the first attempts to improve the user privacy and browsing performance against the most prevalent problem surfing the web at the time; the increasing amount of intrusive and malicious advertisements spreading over the Internet. 

With time, the Internet has grown and some alternatives, often more specific, like JavaScript blockers, Flash blockers or third party blockers, appeared to address the same problem. Unfortunately, unlike general content-filtering tools, using them often affected the usability of websites, making some of them even inaccessible. For that reason, content-blockers have become the \textit{de facto} solution to the advertisement and tracking ecosystem problem.

As recently shown in \cite{characterizing_plugins}, one of the main motivations to use a content-filtering system is to improve the browsing experience. Almost one third of the users that install a blocking system use it to increase the browsing performance. The main reason behind this is that blocking advertisements and tracking systems is believed to significantly reduce the bandwidth used and improve the website loading time.

In this paper, we present a comprehensive study of the advertisement and tracking ecosystem with special emphasis on the performance gains resulting from using content-blockers when surfing the Internet. 
Some previous works have already analyzed and compared the effectiveness of existing content-filtering systems in terms of blocking performance \cite{adblocker_do, tracker_blocker_comparison}. However, to the best of our knowledge, this is the first work to compare content-blockers in terms of bandwidth and latency improvements with a large and diverse set of websites.


We measure the latency and bandwidth usage when visiting the top 100K sites according to the Alexa list \cite{alexa_list}. We compare three different blocking approaches; \textit{advertisement blocking}, \textit{tracker blocking} and \textit{generic content blocking}. For this purpose, we developed a highly parallel system that loads every website using one of the most relevant content-blockers of each mentioned group and compares their performance.

We found that, although there is some gain from using a content-filtering plugin in terms of bandwidth, the results do not translate directly to latency gains and, in some cases, there could even be a penalty to be paid. This is the case for one of the plugins studied, especially for low-latency websites.

The rest of the paper is organized as follows. Section \ref{bg} provides the background on the topic as well as the related work. Section \ref{methodology} introduces the methodology used to collect the data and perform the experiments. Section \ref{results} presents the obtained results in terms of bandwidth and latency. Lastly, Section \ref{conclusions} concludes the paper and discusses the future work.

\section{Background and Related Work}
\label{bg}
\subsection{Background}
\label{background}
Advertisements have been used as a way to pay for online expenses as well as for getting benefits from online resources since the foundation of the Internet. At first, the advertisement ecosystem was very simple and owners tried to add some advertisements to monetize each site. Soon, the difficulty to attract user attention became relevant within advertisers. In \cite{advertising_puzzle} there is one of the first studies where that difficulty is stated. To be able to stand out in such an environment companies started to modify their advertisements with sounds and vivid colors, using pop-ups and other intrusive methods to reach their users. Lohtia et al. \cite{impact_of_banners} studied the impact of those methods on advertisement performance and user perception. Such an advertisement environment became a nuisance for some of the users, being the main motivation for the creation of the first \textbf{adblockers}, a browser content-filtering plugin that tries to block all the advertisement being shown in a website.

With the evolution of the Internet and the languages used to develop it, companies started to use profiling mechanisms (e.g. HTTP cookies) over their users to perform targeted advertisement campaigns and to support their services. These primal web tracking techniques were more or less harmless and very simple to avoid or at least diminish their impact. Nowadays companies use much more complex techniques (e.g. Fingerprints, Flash cookies), collecting data not only from their own sites but from other apparently non-related websites (third-party trackers) to refine to the extreme the information collected about the opinions and preferences of their users. In \cite{survey_valentin} Bujlow et al. show a summary of the different mechanisms that can be used to track users and correlate their information. 

Online tracking has raised serious privacy and surveillance concerns. Consequently, a new method of content-filtering appeared; \textbf{tracking blockers}. These plugins try to do the same that adblockers do with advertisements but with the tracking systems included in websites. Usually both of them, \textit{adblockers} and \textit{tracking blockers}, use custom databases or filter lists to distinguish between safe and non-safe content. Two of the most important filter lists are EasyList \cite{easy_list} and EasyPrivacy \cite{easy_privacy}, each one of them dedicated to block advertisements and tracking methods respectively. These lists are maintained by the community and used by some of the most popular content-filtering extensions on the market

Since the system for blocking trackers and advertisements are very similar, adblockers started to introduce at some level the possibility to also block web tracking mechanisms. On top of that, a new type of content-filtering emerged; the generic \textbf{content blocker}. This blocker genre tries to intercept not only advertisements and trackers, but other unneeded resources or security threats that could be detected in the website loading process, like for instance Cross-Site Scripting (XSS) resource loading.

On the other hand, many online publishers have become part of the \textit{acceptable ads program} \cite{acceptable_ads_program}, which allows them to avoid being blocked by some of the adblockers if their advertisements follow specific guidelines that ensures them to be less intrusive. 

\subsection{Related work}

Krishnamurthy et al. \cite{privacy_loss} was one of the first to study the impact of adblockers and other privacy protection mechanisms in web browsing. The study included measurements of the impact on page functionality and quality when using such protection mechanisms.

In \cite{adblocker_do} Wills et al. makes a comparison between different adblockers and tracker-blockers, studying the success rate of their methods to detect and block the content of different third party trackers. Traverso et al. \cite{tracker_blocker_comparison} study the behave of seven different content-filtering plugins within a set of 100 specific websites classified in different popular categories. Studies both, the tracker blocking success rate and the performance gains.

Pujol et al. \cite{adblock_in_the_wild} studied the usage of adblockers in an European ISP and found that 22\% of the most active users used an adblocker. Malloy et al. \cite{adblock_impact} measured the percentage of adblockers users in the U.S. and infered that about 18\% was using it.

Shiller et al. \cite{blocking_effect} tried to measure the impact of advertisement blocking on the quality of content created, and revenues of online publishers. The constantly increasing adoption rate and lose of revenues made online publishers to start using different techniques to try to bypass adblockers. In \cite{ad_wars} Iqbal et al. studied the anti-adblocking ecosystem and found it to be continuously increasing.

In \cite{internet_jones} Lerner et al. studied the prevalence of tracking methods, used for targeted advertising and other purposes, over the Internet. Also studied it's increase over time using the Wayback Machine. They found that almost 70\% of websites use some type of tracker, enforcing the use of privacy protections like trackers blockers.

From the user's perspective there are different reasons to use a content-filtering plugin. Mathur et al. \cite{characterizing_plugins} studied the different content-filtering group adoption (\textit{adblocker}, \textit{tracker blocker} or \textit{general content blocker}) as well as the different and common motivations to use all of them. 

They found \textit{adblockers} to be the most prevalent blocker system (51.2\%), with \textit{general content blocker} next (20.5\%) and \textit{tracker blocker} in the last position (8.4\%). The primary reason to use an \textit{adblocker} or a \textit{general content blocker} (the two most adopted solutions) was to improve the user experience (85-89\%), and between the most common given motivations was \textbf{speedup loading times} with a 33.1\% of acceptance. Other reasons were to improve privacy (mainly by means of \textit{tracker blockers}) and only a small percentage used blockers to improve security. 

This work tries to clarify if there is a relation between content-filtering usage and an increase in browsing performance, as one third of the plugin users seems to think. To the best of our knowledge the only publication very related to our study is the one performed by Traverso et al. \cite{tracker_blocker_comparison} using a population of 100 websites with most of them pertaining to the same country. Until now there is no prior publication focused on studying content-filtering usage from a performance perspective in a large-scale environment.

\section{Methodology and evaluation}
\label{methodology}
To explore the performance gain introduced by content-filtering plugins we will explore both of the principal parameters that can represent an increase in performance; bandwidth and latency. To be able to generalize our results we have to accomplish several conditions. 

\begin{enumerate}[label=\Alph*]
    \item \textbf{Dataset}: The number of websites to browse has to be big enough to be able to extrapolate the results to a bigger population, and the selected websites should be representative enough of an usual browsing session.
    \item \textbf{Browsing experience}: The website has to be loaded exactly as it would be done if accessed by a real user. This will avoid websites to detect automation scripts that would possibly make them not to load all the resources.
    \item \textbf{Loading interval restrictions}: The same website loading process using different content-filtering plugins should be performed in a small time window to avoid whenever possible different temporal conditions (e.g. rush hours, periodic maintenances, etc).
    \item \textbf{Experiments repetition}: The number of repetitions has to be higher enough to be able to discard possible outliers, especially in latency experiments, where network conditions can change from one execution to another.
    \item \textbf{Browsing completion}: We have to measure all the resources being loaded, even third party ones or resources loaded dynamically and not included in the website code.
\end{enumerate}

\subsection{Dataset}
\label{dataset}
For the dataset we decided to use the top 100K most popular websites according to the Alexa list \cite{alexa_list}. Scraping the most popular websites will give us a good approximation of what a real user would access. Alexa's list is part of the Amazon ecosystem and has been used in several publications in the past (e.g. \cite{privacy_loss}, \cite{one_million}). Note that we only examine the homepages of each one of the websites included in the list but none of the links and external resources available within them.

\subsection{Browsing Experience}
\label{browsing}
We developed our own highly parallelized system making use of \textit{Selenium} \cite{selenium} for the automation process. This allows us to run real browsers such as \textit{Mozilla Firefox} or \textit{Google Chrome} that will present identical results a user would normally get. This also permits us to customize all the experiments using browser settings and network headers. Selenium has been used in the past for automation of research experiments in several different topics (e.g. \cite{one_million}, \cite{tacking_free}).

\begin{table}
  \caption{Content-filtering plugins comparison}
  \label{tab:blockers}
  \resizebox{0.485\textwidth}{!}{
  \begin{tabular}{|l|c|c|c|}
    \hline
    \textbf{}&\textbf{AdBlock Plus}&\textbf{Ghostery}&\textbf{uBlock Origin}\\
    \hline
    \textbf{Ads Blocking}&$\fullmoon$&$\fullmoon$&$\fullmoon$\\
    \hline
    \textbf{Acceptable Ads}&$\fullmoon$&$\bigtriangleup$&\---\\
    \hline
    \textbf{Track Blocking}&\---&$\fullmoon$&$\fullmoon$\\
    \hline
    \textbf{EasyList}&$\fullmoon$&\---&$\fullmoon$\\
    \hline
    \textbf{EasyPrivacy}&Available&\---&$\fullmoon$\\
    \hline
    \textbf{Others}&\---&Private Database&Additional lists\\
    \hline
  \end{tabular}}
\end{table}

The experiments are done browsing the 100K commented websites using one plugin from each one of the groups presented in Section \ref{background}; \textbf{AdBlock Plus} as an \textit{adblocker}, \textbf{Ghostery} as a \textit{tracker blocker} and \textbf{uBlock Origin} as a generic \textit{content blocker}. All of them have versions available for the most common internet browsers like \textit{Google Chrome} or \textit{Mozilla Firefox}. A summary of the differences between them is shown in table \ref{tab:blockers}.

We will use as well a vanilla browser as a base to compare all the observed information. We decided to use \textit{Chromium}, the open-source version of Google's Chrome, as the main tool to scrap the data. We made some tests using \textit{Mozilla Firefox} with a small dataset to compare the results and observe if there was any substantial difference. We found that performance is equivalent using both of them. Thus, we opted for Chromium as it permits us to load in an easy and straightforward way the needed plugins as well as to get information about the network communications being performed.

Regarding the plugins, we decided to use the default settings for all of them, as usually users do not change them after installation. Note that in \textit{AdBlock Plus} the \textit{acceptable ads program} list is enabled by default and we leave it this way. The only modification introduced is adding EasyPrivacy subscription to \textit{AdBlock Plus}. We have two main reasons for those decisions. First, we want to test the three plugins in equal conditions, blocking both advertisements and web tracking mechanisms. Secondly, as observed in \cite{adblock_in_the_wild}, \textit{AdBlock Plus} users usually activate EasyPrivacy list even if it is disabled by default, but do not disable the \textit{acceptable ads program}, probably because they are more interested in blocking unneeded content than in privacy. On the other hand, Ghostery does not use the \textit{acceptable ads program} list, but it has its own monetizing program introducing advertisements instead of letting them pass through.

\subsection{Loading interval restrictions}
\label{interval}
Specially for latency experiments we have to assure that the measures are taken for all the plugins in a short period of time. To accomplish this, our developed system opens the same website with 4 different browsers in parallel, three of them loading the corresponding plugin and the last one with a vanilla browser.

\subsection{Experiments repetition}
\label{repetitions}
For bandwidth experiments it is enough to scrap the resources being loaded by the browser once for each of the content-filtering plugins, and once more using a vanilla browser without any of them. This gives us enough information to compare the bandwidth improvement using each of the plugins.

In comparison, for the latency experiments the same website is opened in parallel by the four browsers 5 times consecutively, having a timeout setting of 30 seconds, usually bigger enough to load all the resources. Between each one of the repetitions the browsing cache as well as the cookies are deleted to obtain a clear browsing experience. On top of that, if one of the browsers is unable to get 5 measures of the same website, all the website measures are discarded (even measures gotten by the other browsers), as it is not possible to compare the results with the other content-filtering plugins. With this system, from the initial population of 100k websites we ended up scrapping successfully a total of 83.374 websites. All the measures were taken in the period May-July of 2019.

\subsection{Browsing completion}
\label{completion}
Nowadays most of the websites include at some level script files that loads resources dynamically. Moreover, almost all of them are obfuscated or minifyed; a process that tries to reduce the size of the script files and improve the loading time by removing white-spaces, break-lines and shortening the names of the variables. These two facts make impracticable to get a list of the resources being downloaded exploring the code in the traditional way.

\begin{table}
  \caption{Performance Timing API}
  \label{tab:timing}
  \resizebox{0.485\textwidth}{!}{
  \small
  \begin{tabular}{|l|c|}
    \hline
    \multicolumn{1}{|c|}{\textbf{Event}} & \textbf{Stage} \\ \hline
    startTime & \multirow{3}{*}{Prompt for unload} \\ \cline{1-1}
    unloadEventStart &  \\ \cline{1-1}
    unloadEventEnd &  \\ \hline
    redirectStart & \multirow{2}{*}{Redirect} \\ \cline{1-1}
    redirectEnd &  \\ \hline
    fetchStart & AppCache \\ \hline
    domainLookupStart & \multirow{2}{*}{DNS} \\ \cline{1-1}
    domainLookupEnd &  \\ \hline
    connectStart & \multirow{3}{*}{TCP} \\ \cline{1-1}
    secureConnectionStart &  \\ \cline{1-1}
    connectEnd &  \\ \hline
    requestStart & \multirow{2}{*}{Request} \\ \cline{1-1}
    responseStart &  \\ \hline
    responseEnd & Response \\ \hline
    domInteractive & \multirow{4}{*}{Processing} \\ \cline{1-1}
    domContentLoadedEventStart\hspace{2.8cm} &  \\ \cline{1-1}
    domContentLoadedEventEnd &  \\ \cline{1-1}
    domComplete &  \\ \hline
    loadEventStart & \multirow{2}{*}{Load} \\ \cline{1-1}
    loadEventEnd &  \\ \hline
  \end{tabular}}
  \vspace{-0.4cm}
\end{table}

To get the resource bandwidth information we make use of Google's \textit{DevTools Protocol} to get access to all the network communications being executed by the browser. We do it by enabling the logging capabilities of the browser and extracting the information directly from the network logs in real time. This way we ensure that all the traffic is being collected even if redirections occurs. If a redirection occurs when a resource is being loaded we add the bandwidth used for that redirection to the total bandwidth needed for loading that specific resource.

As for the latency metrics, we use the \textit{Performance Timing} API \cite{performance_timing} defined by the W3C and supported by the majority of the current browsers to extract the information of all the loading events each website produces. In Table \ref{tab:timing} it is shown the events produced by the \textit{Performance Timing} API in execution order when a website loads. Using the time difference between those events we can compute different measures, as for instance the total page load time or the \textit{Above The Fold} latency (latency to load only the portion of the website that is initially seen inside the browser). Getting the total load time would also take into account the time to unload the current website as well as the time spent in redirections. As we only want to account the time difference between loading the website with and without content-filtering plugin, we will focusing our experiments computing the time between the \textit{requestStart} event and the \textit{domComplete} event, that gives us the total latency from the point when the browser makes the actual request until it loads all the needed resources.

\begin{figure}
    \centering
	\includegraphics[width=0.475\textwidth]{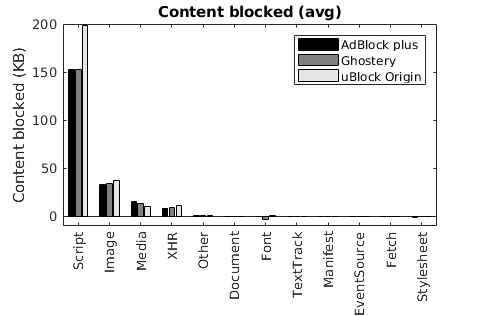}
	\label{fig:content_blocked}
	\caption{Average content blocked}
  \vspace{-0.4cm}
\end{figure}

\begin{figure*}[!t]
  \minipage{0.475\textwidth}
  \includegraphics[width=\linewidth]{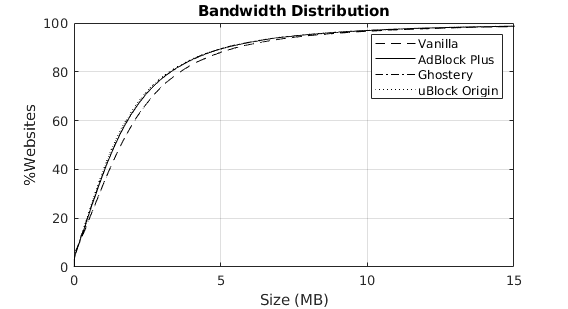}
	\label{fig:adblock_comparison}
	\caption{Bandwidth comparison}
  \endminipage\hfill
  \minipage{0.475\textwidth}
  \centering
  \includegraphics[width=\linewidth]{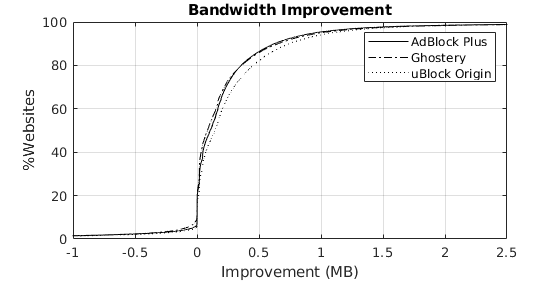}
	\label{fig:bw_improvement}
	\caption{Bandwidth improvement}
  \endminipage
\end{figure*}

\begin{figure}
    \centering
	\includegraphics[width=0.475\textwidth]{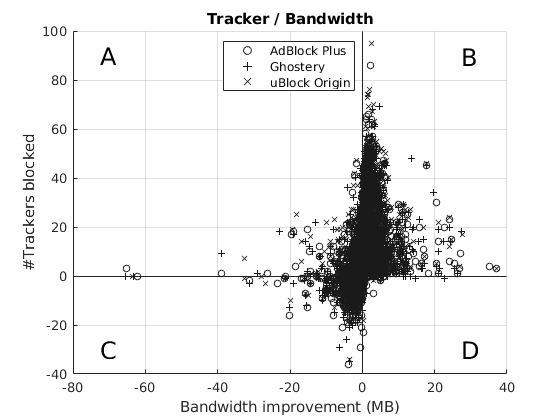}
	\label{fig:trackers_bandwidth}
	\caption{Trackers / Bandwidth}
\end{figure}

As commented on Section \ref{repetitions}, the latencies have been acquired a total of 5 times for each website and for each of the plugins used (\textit{Adblock Plus}, \textit{uBlock Origin} and \textit{Ghostery}) as well as for the vanilla browser for comparison reasons. All websites that have less than 5 observations for any of the mentioned configurations have been excluded. Furthermore, to avoid influence of external circumstances we have discarded outliers using the \textbf{interquartile range}. Then, we have computed the average for the rest of the observations to obtain an approximation of the loading latency.

\section{Results}
\label{results}
\subsection{Bandwith comparison}
\label{bandwidth}
One of the terms that can be used to compare a website loading performance is the bandwidth needed to load all the resources included in it. A vanilla browser will load all of them while a content-filtering included browser will only load the ones not being blocked. Intuitively, it should represent a big difference in terms of data accessed and could be of especial importance for mobile devices, where the difference could lead to extra expenses. 

Figure 1 shows the average of blocked resources per site classified by type. Here \textit{uBlock Origin} presents a slight gain margin against \textit{Ghostery} and \textit{AdBlock Plus}. This difference is not unexpected and should be due to the extra contents being blocked by a generic content blocker, as well as for the inclusion of some advertisements by the other two contestants.

In Figure 2 we show the comparison of the total bandwidth needed for loading the top 100k most popular websites with each one of the content-blockers. At first sight it can be seen that it exists a difference but it is very subtle. Figure 3 represents the bandwidth improvement over the total size of the website. In both figures the three plugins selected for the experiments follow the same progression, showing that there is almost no real difference between selecting any of them in terms of bandwidth gain.

Figure 4 shows the relation between the number of trackers being blocked by each plugin and the bandwidth gain introduced. The results can be divided using the coordinates axes in four quadrants named from \textbf{A} to \textbf{D} in the graph. In Table \ref{tab:track_band} it is shown the percentage of websites included in each quadrant for all the plugins. The results are very similar for the three content-blockers compared. 

The majority of the observations are situated in the quadrant \textbf{B}, with positive values for both axis. This is the expected situation, meaning that there is a relation between blocking trackers and improving the bandwidth usage. Surprisingly, the relation is not linear, and blocking more trackers does not necessarily imply diminishing the bandwidth used. 

The graph also shows that the quadrant \textbf{C}, the opposite case, is not rare as there is a not negligible percentage of websites included in it. This fact also implies a relation between trackers and bandwidth; there are new trackers being loaded that have repercussions as a bandwidth penalty. The percentage is small ($\approx4\%$) and is mostly due to the dynamic way of the website loading process. Each time a website loads, the content slightly varies including new information depending on the third party trackers loaded. Nevertheless, there are some cases where the difference is big enough to be considered relevant.

For the remaining two quadrants, \textbf{D} with only a few observations is very intuitive. It would be very uncommon to have bandwidth gains if the number of third party trackers increases. Unexpectedly, the number of observations in quadrant \textbf{A}, where blocking trackers has an negative impact in the total bandwidth used to load the website, is relevant (3\% to 6\%). This unnatural behaviour and the special cases in the quadrant C remains to be studied in the future work.

\begin{table}
  \caption{Tracker / Bandwidth distribution}
  \label{tab:track_band}
  \resizebox{0.485\textwidth}{!}{
  \small
  \begin{tabular}{|c|c|c|c|}
  \hline
  \textbf{Quadrant} & \textbf{AdBlock Plus} & \textbf{Ghostery} & \textbf{uBlock Origin} \\ \hline
  \textbf{A} & 3.81\% & 6.21\% & 3.45\% \\ \hline
  \textbf{B} & 84.73\% & 81.62\% & 86.27\% \\ \hline
  \textbf{C} & 3.52\% & 4.4\% & 2.63\% \\ \hline
  \textbf{D} & 0.17\% & 0.15\% & 0.08\% \\ \hline
  \textbf{Origin (0,0)} & 7.76\% & 7.6\% & 7.56\% \\ \hline
  \end{tabular}}
  \vspace{-0.4cm}
\end{table}

\begin{figure*}[!t]
  \minipage{0.475\textwidth}
  \includegraphics[width=\linewidth]{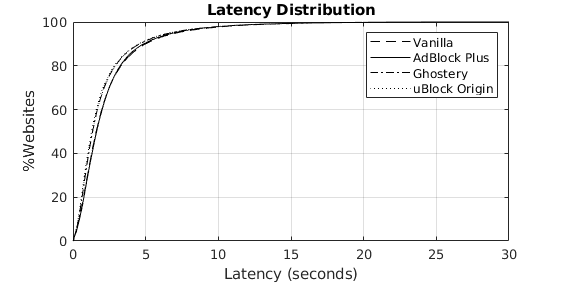}
  \label{fig:ublock_lat_comparison}
  \caption{Latency comparison}
  \endminipage\hfill
  \minipage{0.475\textwidth}
  \centering
  \includegraphics[width=\linewidth]{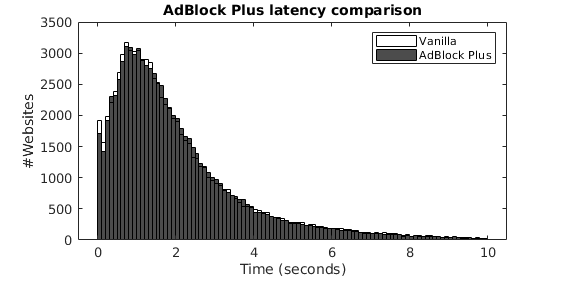}
  \label{fig:adblock_lat_improvement}
  \caption{Adblock latency improvement}
  \endminipage
  \vspace{-0.4cm}
\end{figure*}

\begin{figure}
  \centering
  \includegraphics[width=\linewidth]{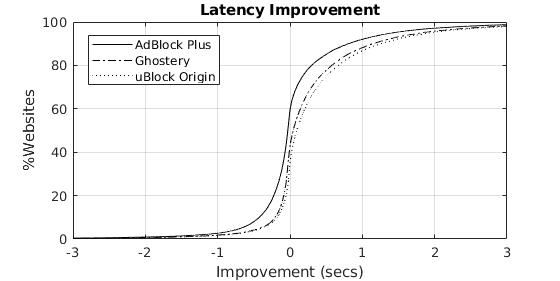}
  \label{fig:lat_improvement}
  \caption{latency improvement}
  \vspace{-0.4cm}
\end{figure}

\subsection{Latency comparison}

The second parameter that can be used to measure a browsing experience improvement is the latency needed to load websites. If introducing a content-filtering plugin represents an important decrease in latency, their users can take profit not only from an improvement in privacy and security, but in browsing speed too.

The latency distribution achieved with our population of 100k websites is shown in Figure 5. Approximately 90\% of the websites have a latency lower than 5 seconds, being this range the one with the most benefit of using a content-blocker. Nevertheless, as with the bandwidth experiments, the difference is very small, with 318 milliseconds of average gain.

Note that, even being almost unnoticeable, the curve presented by \textit{AdBlock Plus} is almost exact to the one presented by the vanilla browser. To see the differences we show in Figure 6 the histogram comparison between both of them. If we take a closer look at the range lower than 2 seconds, the quantity of domains that load faster without the adblocker is higher than with it. In terms of performance we can conclude that using \textit{AdBlock Plus} for browsing very fast-loading websites introduces a penalty not seen in vanilla browsers. As \textit{AdBlock Plus} does not inject any kind of traffic to the loading process of a website, like for instance \textit{Ghostery} does with their monetizing program, we conjecture that it is related to the overhead introduced by the plugin itself to check if online resources must be blocked or not.

Figure 7 plots the CDF with the latency improvement in seconds for each of the three content-filtering plugins. Here, unlike in the bandwidth experiments, we can clearly see a difference between them. In terms of percentage \textit{AdBlock Plus} presents a penalty of about minus 10\% on average while \textit{Ghostery} and \textit{uBlock Origin} presents an improvement of 3.1\% and 9.8\% respectively. This is the gain range factor that we can expect using a content-blocker.

\section{Conclusions and Future Work}
\label{conclusions}
In this work, we presented the results of a comprehensive measurement study that analyzes the actual performance gains of using content-filtering plugins in terms of latency and bandwidth savings. Contrary to common belief, our results show that the improvements in terms of bandwidth are small, while speed ups in page loading times are almost negligible. 

Regarding the relative performance among the analyzed tools, we did not observe significant differences when measuring the bandwidth, but we discovered some interesting differences in terms of latency, where \textit{AdBlock Plus} performed worse than \textit{Ghostery} and \textit{uBlock Origin}, even introducing a penalty for fast-loading websites.

Based on our results, we can conclude that the use of content-filtering plugins for performance reasons can still be useful for slow connections, where bandwidth gains of about $10\%$ can make a difference. On the contrary, for normal connections where the throughput is not a barrier and, within the current Internet context, where loading times are below 5 seconds, a latency gain of 318 milliseconds does not seem an important performance improvement.

As a future work, we plan to analyze in more detail the bandwidth penalty observed in Section \ref{bandwidth} when blocking some third party trackers. We also intend to characterize the latency and access times depending on the source and destination of the connections, looking for differences between countries and content type.

\section{Acknowledgments}
This work was supported by the Spanish MINECO under contract TEC2017-90034-C2-1-R (ALLIANCE).

\end{document}